# A molecular equation of state for alcohols which includes steric hindrance in hydrogen bonding

Bennett D. Marshall

*ExxonMobil Research and Engineering, 22777 Springwoods Village Parkway, Spring TX 77389 USA*

## Abstract

In this paper we develop the first equation of state for alcohol containing mixtures which includes the effect of steric hindrance between the two electron lone pair hydrogen bond acceptor sites on the alcohol's hydroxyl oxygen. The theory is derived for multi-component mixtures within Wertheim's multi-density statistical mechanics in a second order perturbation theory. The accuracy of the new approach is demonstrated by application to pure methanol and ethanol and binary ethanol / water mixtures. It is demonstrated that the new approach gives a substantial improvement in the prediction of hydrogen bonding structure of both pure alcohol and alcohol/water mixtures as compared to conventional approaches which do not include steric effects between the alcohol association sites. Finally, it is demonstrated that the inclusion of steric effects allows for more accurate binary phase equilibria and heats of mixing prediction with water.

Bennettd1980@gmail.com

**I: Introduction**

Much of current thermodynamics research is focused on molecular simulations: new algorithms, force fields, applications, etc.… Indeed, molecular simulations allow for the "exact" solution of the thermodynamics / fluid structure for a given force field. However, even with the advances in molecular simulation, it is the general mixture equation of state (EoS) which forms the backbone of commercial thermodynamics packages used to predict fluid phase equilibria. EoS can be used to quickly estimate the phase equilibria of multi-component mixtures, which makes them amiable for implementation into process simulators.

The predictive ability of an EoS often relies on a proper accounting of the physics of the relevant intermolecular interactions. Perturbation theory provides a relatively simple and accurate framework in which to develop EoS. Weeks, Chandler, Anderson[1] and Barker, Hendersen[2] where the first widely applied perturbation theories for spherically symmetric fluids. Extension of Perturbation theory to hydrogen bonding (associating) fluids is challenging due to the strength, anisotropy and limited valence of the hydrogen bond.

Andersen[3,4], Dahl and Andersen[5] and Chandler and Pratt[6] where the first to develop statistical mechanical formalisms which could be used to develop perturbation theories for associating fluids. However, these approaches have not been widely applied because the final theoretical results are in the form of cluster expansions. In the 1980's Wertheim[7–10] developed a new multi-density form of statistical mechanics, where each hydrogen bonding state of a molecule is assigned a separate density. Taking this approach, Wertheim was able to develop a rigorous statistical mechanics formalism for associating fluids. The beauty of Wertheim's approach is that general solutions to the theory can be obtained in perturbation theory (TPT). Chapman[11] obtained a general solution to TPT in first order (TPT1) for a multi-component mixture where each molecule can have an arbitrary number and functionality of association sites. The simplicity and generality of this TPT1 solution has allowed for its wide industrial and academic application in the form of the statistical associating fluid theory[12–16] (SAFT) class of EoS as well as the Cubic-Plus-Association[17] EoS.

While widely applied, TPT1 has many limitations. First order perturbation theory assumes that each association bond in a cluster is independent of the other association bonds. This allows for a simple general solution, but neglects effects such as steric hindrance[18,19], ring formation[20,21], double bonding of molecules[22,23] and association sites which can receive multiple association bonds[24,25]. To include these effects, one must go to higher order in perturbation. TPT2 allows for the interaction of two association bonds in a cluster, TPT3 allows for the simultaneous interaction of 3 association bonds etc.…

Alcohols are a common class of associating molecules. Each alcohol has two hydrogen bond acceptor sites (oxygen lone pairs) and one hydrogen bond donor (hydroxyl hydrogen). This defines a 3C association scheme.[26] There has been much debate in the literature on whether alcohols should be modeled with two (one donor and one acceptor in a 2B scheme) or three association sites.[27,28] Recently, Fouad *et al.*[28] demonstrated that within the polar PC-SAFT equation of state, the 2B scheme gave better agreement with hydrogen bonding distributions in pure alcohols. However, it was simultaneously shown that the 3C model gave better agreement for hydrogen bonding distributions for alcohol-water mixtures, as well as well as improved phase behavior predictions with water.

The results of Fouad *et al.*[28] point to a deficiency in the application of TPT1 to alcohols. As TPT1 does not include steric effects, it will not account for the fact that bonding at one of the oxygen acceptor sites may partially block accessibility to the other. This would lead to an overprediction in the fraction of ethanol molecules which are bonded at both acceptor sites. Hence, a 2B model, which rejects second acceptor site completely, gave better results than the 3C model. However, when inserting an alcohol molecule in an aqueous phase, it may become bonded at all three sites. What is needed is a theoretical approach which accounts for the steric hindrance between the two oxygen sites in a 3C model. This will be the subject of this paper.

In this paper we develop the first general multi-component solution of TPT2 which allows for the incorporation of steric effects. Each species can have an arbitrary number and functionality of association sites; however, for tractability we only allow one pair of sites per molecule to become sterically coupled. We then specialize this solution to mixtures which contain alcohols modelled with a 3C association scheme. We then incorporate the new perturbation theory into the perturbed chain statistical associating fluid theory (PC-SAFT) EoS and use it to study the hydrogen bonding and phase behavior of pure alcohol and alcohol-water binary mixtures. In this paper we focus on the alcohols methanol and ethanol. We demonstrate that the inclusion of steric hindrance through TPT2 gives a significantly improved EoS as compared to TPT1.

## II: Thermodynamic perturbation theory

In this section we extend thermodynamic perturbation theory (TPT) to account for steric hindrance between association sites in a multi-component fluid. We consider a mixture of $N$ molecules consisting of $n$ separate species of number density $\rho^{(k)}$. Each species contains a set of $\Gamma^{(k)} = \{A, B, C,...,G\}$ association sites, where the capitals letters represent distinct association sites. While each species can have any number and type of association sites, we restrict the theory such that only a single pair of association sites per species can sterically hinder each other. For molecules with a 3C association scheme, we call this approach the 3C-SH association model. Figure 1 gives a model representation of a 3C-SH alcohol model. There is one donor hydrogen labeled $H$ and two acceptor oxygens lone pairs which we label $O_1$ and $O_2$. However, only the two oxygen sites exhibit steric effects.

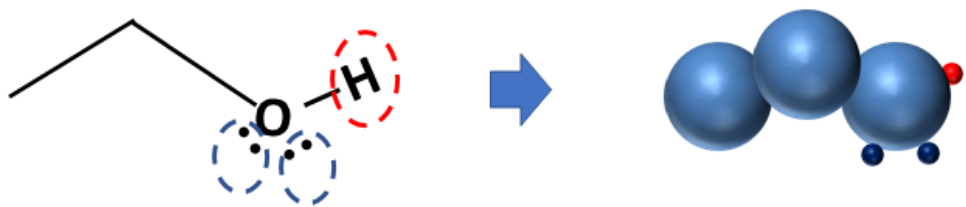

**Figure 1:** Example representation of ethanol using a 3-C association scheme. The two lone pairs of electrons are treated as acceptor sites (blue) and the hydrogen is a donor site (red)

The potential of interaction between a molecule 1 of species $k$ and a molecule 2 of species $j$ is given by[9]

$$\varphi^{(k,j)}(12) = \varphi_{hs}^{(k,j)}(r_{12}) + \sum_{A \in \Gamma^{(k)}} \sum_{B \in \Gamma^{(j)}} \varphi_{AB}^{(k,j)}(12) \qquad (1)$$

The distance between the centers of the molecules is $r_{12}$ and the notation (1) represents the position and orientation of molecule 1. The term $\varphi_{hs}$ is the pair potential of the spherically symmetric hard sphere reference fluid and $\varphi_{AB}^{(k,j)}$ is the potential of interaction between site *A* on species *k* and site *B* on species *j*.

The theory is developed in Wertheim's multi-density formalism.[9] In this approach each bonding state of a molecule is treated as a distinct species and assigned a density $\rho_\alpha^{(k)}$, where α is the set of bonded sites. Hence, $\rho_o^{(k)}$ is the monomer density of species *k*. To aid in the topological reduction from fugacity to density graphs, Wertheim defines a set of density parameters

$$\sigma_\alpha^{(k)} = \sum_{\gamma \subset \alpha} \rho_\alpha^{(k)} \tag{2}$$

where $\sigma_o^{(k)} = \rho_o^{(k)}$ and $\sigma_\Gamma^{(k)} = \rho^{(k)}$. The total Helmholtz free energy is given by[10]

$$\frac{A - A_{hs}}{V k_B T} = \sum_k \left( \rho^{(k)} \ln\left(\frac{\rho_o^{(k)}}{\rho^{(k)}}\right) + Q^{(k)} + \rho^{(k)} \right) - \frac{\Delta c^{(o)}}{V} \tag{3}$$

where $A_{hs}$ is the free energy of the hard sphere reference fluid, *V* is the system volume and *T* is the absolute temperature.

Equation (3) is mathematically exact. The fundamental graph sum $\Delta c^{(o)}$ is an infinite series of integrals which encodes all association interactions between molecules. In thermodynamic perturbation theory (TPT) all contributions to $\Delta c^{(o)}$ which contain more than a single associated cluster are neglected. This allows for the summation of $\Delta c^{(o)}$ in terms of reference system correlation functions. TPT assumes that the structure of the fluid is unchanged due to association. Within perturbation theory, $\Delta c^{(o)}$ is then ordered in terms of irreducible clusters which contain 1, 2, 3 etc… association bonds

$$\Delta c^{(o)} = \sum_j \Delta c_j \tag{4}$$

where $\Delta c_j$ contains irreducible contributions from cluster integrals with $k$ association bonds. In first order perturbation theory (TPT1) Eq. (4) is truncated at $j = 1$, second order TPT2 at $j = 2$ etc… At the TPT1 level, Eq. (4) only includes contributions with a single irreducible bond. All reducible clusters can be created from the TPT1 result, however there is no steric hindrance between association sites. For a multi component mixture, where each species can have an arbitrary number and functionality of association sites, the TPT1 contribution is given by

$$\frac{\Delta c_1}{V} = \frac{1}{2}\sum_{k=1}^{n}\sum_{j=1}^{n}\sum_{A\in\Gamma^{(k)}}\sum_{B\in\Gamma^{(j)}} \sigma_{\Gamma^{(k)}-A}^{(k)} \sigma_{\Gamma^{(j)}-B}^{(j)} \Delta_{AB}^{(k,j)} \tag{5}$$

Where

$$\Delta_{AB}^{(k,j)} = \frac{1}{8\pi^2}\int f_{AB}^{(k,j)}(12) g_{hs}^{(k,j)}(r_{12}) d(12) \tag{6}$$

The association Mayer function is given by

$$f_{AB}^{(k,j)}(12) = \exp\left(-\frac{\varphi_{AB}^{(k,j)}(12)}{k_b T}\right) - 1 \tag{7}$$

and $g_{hs}^{(k,j)}(r_{12})$ is the mixture pair correlation function of the hard sphere reference system.

Including the TPT2 term includes irreducible graphs which contain clusters with two association bonds. This is the minimum level of theory to incorporate steric effects. Extending the second order approach of Wertheim[29] to the case of a multi-component fluid with and arbitrary number of association sites

$$\frac{\Delta c_2}{V} = \frac{1}{2}\sum_{k=1}^{n}\sum_{j=1}^{n}\sum_{i=1}^{n}\sum_{A\in\Gamma^{(k)}}\sum_{B\in\Gamma^{(j)}}\sum_{C\in\Gamma^{(i)}}\sum_{D\in\Gamma^{(i)}} \sigma_{\Gamma^{(k)}-A}^{(k)} \sigma_{\Gamma^{(j)}-B}^{(j)} \sigma_{\Gamma^{(i)}-CD}^{(i)} \Delta_{ACDB}^{(k,i,j)} \tag{8}$$

with

$$\Delta^{(k,i,j)}_{ACDB} = \frac{1}{64\pi^4}\int f^{(k,i)}_{AC}(12) f^{(i,j)}_{DB}(23) g^{(k,i)}_{hs}(r_{12}) g^{(i,j)}_{hs}(r_{23}) \times$$
$$\left(\frac{g^{(k,i,j)}_{hs}(r_{12},r_{23},r_{13})}{g^{(k,i)}_{hs}(r_{12}) g^{(i,j)}_{hs}(r_{23})} - 1\right) d(2)d(3) \qquad (9)$$

Where $g^{(k,i,j)}_{hs}(r_{12},r_{23},r_{13})$ is the triplet correlation function of the reference fluid.

Equation (9) completes the definition of the graph sum $\Delta c^{(o)}$. The last term to consider in Eq. (3) is $Q^{(k)}$

$$Q^{(k)} = -\rho^{(k)} + \sum_{\substack{\gamma \subset \Gamma^{(k)} \\ \gamma \neq \emptyset}} c^{(k)}_\gamma \sigma^{(k)}_{\Gamma^{(k)}-\gamma} \qquad (10)$$

The functions $c^{(k)}_\gamma$ are generated from the graph sum $\Delta c^{(o)}$ according to the relation

$$c^{(k)}_\gamma = \frac{\partial}{\partial \sigma^{(k)}_{\Gamma-\gamma}} \frac{\Delta c^{(o)}}{V}; \quad \gamma \neq \emptyset \qquad (11)$$

Now we assume that each molecule can have at most one pair of second order sites which sterically hinder each other. We label this set {C,D}. Evaluating Eq. (11) subject to Eqns. (4), (5) and (8)

$$c^{(k)}_A = \sum_{j=1}^n \sum_{B \in \Gamma^{(j)}} \sigma^{(j)}_{\Gamma^{(j)}-B} \Delta^{(k,j)}_{AB} + \frac{1}{2} \sum_{j=1}^n \sum_{i=1}^n \sum_{B \in \Gamma^{(j)}} \sum_{C \in \Gamma^{(i)}} \sum_{D \in \Gamma^{(i)}} \sigma^{(j)}_{\Gamma^{(j)}-B} \sigma^{(i)}_{\Gamma^{(i)}-CD} \left(\Delta^{(k,i,j)}_{ACDB} + \Delta^{(j,i,k)}_{BCDA}\right) \qquad (12)$$

$$c^{(k)}_{EF} = \begin{cases} \frac{1}{2}\sum_{i=1}^n \sum_{j=1}^n \sum_{A \in \Gamma^{(i)}} \sum_{B \in \Gamma^{(j)}} \sigma^{(i)}_{\Gamma^{(i)}-A} \sigma^{(j)}_{\Gamma^{(j)}-B} \left(\Delta^{(i,k,j)}_{ACDB} + \Delta^{(i,k,j)}_{ADCB}\right) & \text{for } EF = CD \\ \\ 0 & \text{otherwise} \end{cases} \qquad (13)$$

$$c^{(k)}_\gamma = 0 \quad \text{for} \quad n(\gamma) > 2 \qquad (14)$$

A key quantity in TPT1 is the fraction molecules not bonded at site A: $X_A^{(k)} = \sigma_{\Gamma^{(k)}-A}^{(k)}/\rho$. In this second order theory we will also require the fraction of molecules not bonded at both sites $C$ and $D$: $X_{CD}^{(k)} = \sigma_{\Gamma^{(k)}-CD}^{(k)}/\rho$.

With the current assumption of only a single pair of second order sites the theory will have similar structure to that of Marshall and Chapman (MC)[21]. Generalizing the MC solution to a multi-component mixture we obtain

$$X_S^{(k)} = \begin{cases} \dfrac{1}{1+c_S^{(k)}} & \text{for } S \neq C \text{ or } D \\[1em] \left(1+c_L^{(k)}\right) X_{CD}^{(k)} & \text{otherwise} \end{cases} \quad (15)$$

In Eq. (15) when $S = C$, $L = D$ and when $S = D$, $L = C$. The fractions $X_{CD}^{(k)}$ are given by

$$X_{CD}^{(k)} = \frac{1}{c_{CD}^{(k)} + \left(1+c_C^{(k)}\right)\left(1+c_D^{(k)}\right)} \quad (16)$$

The monomer fraction is found to be

$$X_o^{(k)} = \frac{X_{CD}^{(k)}}{X_C^{(k)} X_D^{(k)}} \prod_{A \in \Gamma^{(k)}} X_A^{(k)} \quad (17)$$

From these results Eq. (10) can be evaluated as

$$Q^{(k)}/\rho^{(k)} = \sum_{A \in \Gamma^{(k)}} \left(1 - X_A^{(k)}\right) + \frac{X_C^{(k)} X_D^{(k)}}{X_{CD}^{(k)}} - 2 \quad (18)$$

Combining these results, we simplify the free energy in Eq. (3) to

$$\frac{A - A_{hs}}{V k_B T} = \sum_{k=1}^{n} \rho^{(k)} \sum_{A \in \Gamma^{(k)}} \left( \ln X_A^{(k)} - \frac{X_A^{(k)}}{2} + \frac{1}{2} \right) + \sum_{k=1}^{n} \rho^{(k)} \ln\left(\frac{X_{CD}^{(k)}}{X_C^{(k)} X_D^{(k)}}\right) \quad (19)$$

To maintain consistency with the PC-SAFT[13] EoS, equation (6) is evaluated as

$$\Delta_{AB}^{(k,j)} = \sigma_{kj}^3 \kappa_{AB}^{(k,j)} \left( \exp\left( \frac{\varepsilon_{AB}^{(k,j)}}{k_b T} \right) - 1 \right) g_{hs}^{(k,j)} \tag{20}$$

where $\sigma_{kj}$ is the cross-species diameter, $\kappa_{AB}^{(k,j)}$ is the bond volume, $\varepsilon_{AB}^{(k,j)}$ is the association energy and $g_{hs}^{(k,j)}$ is the contact value of the hard sphere pair correlation function evaluated with the Carnahan and Starling[30] EoS. The mixture quantities are evaluated with the following combining rules[13]

$$\sigma_{kj}^3 \kappa_{AB}^{(k,j)} = \sqrt{\sigma_{kk}^3 \kappa_{AB}^{(k,k)} \sigma_{jj}^3 \kappa_{AB}^{(j,j)}} \quad ; \quad \varepsilon_{AB}^{(k,j)} = \frac{\varepsilon_{AB}^{(k,k)} + \varepsilon_{AB}^{(j,j)}}{2} \tag{21}$$

Turning our attention to the second order contribution in Eq. (9), we first approximate the triplet correlation function with the following simple superposition

$$g_{hs}^{(k,i,j)}(r_{12}, r_{23}, r_{13}) = g_{hs}^{(k,i)}(r_{12}) g_{hs}^{(i,j)}(r_{23}) \exp\left( -\frac{\varphi_{hs}^{(k,j)}(r_{13})}{k_b T} \right) \tag{22}$$

The exponential term serves to provide the steric effects between the sites $C$ and $D$ on species $i$. Combining (9) and (22) we obtain

$$\Delta_{ACDB}^{(k,i,j)} = \Delta_{AC}^{(k,i)} \Delta_{DB}^{(i,j)} \left( \Psi_{ACDB}^{(k,i,j)} - 1 \right) \tag{23}$$

Where $\Psi_{ACDB}^{(k,i,j)}$ is the overlap integral defined as the fraction of associated states (where both $C$ and $D$ on $i$ are bonded to site $A$ on $k$ and $B$ on $j$ respectively) which do not lead to overlap. When the sites $CD$ do not sterically hinder each other $\Psi_{ACDB}^{(k,i,j)} = 1$. If there is complete steric hindrance, then the association at one site completely blocks the second site and $\Psi_{ACDB}^{(k,i,j)} = 0$. Mathematically the overlap integral is evaluated by

$$\Psi_{ACDB}^{(k,i,j)} = \frac{\int f_{AC}^{(k,i)}(12) f_{DB}^{(i,j)}(23) \exp\left(-\frac{\varphi_{hs}^{(k,j)}(r_{13})}{k_b T}\right) d(2)d(3)}{\int f_{AC}^{(k,i)}(12) f_{DB}^{(i,j)}(23) d(2)d(3)} \tag{24}$$

In practice we shall assume that the overlap integral of component *i* is independent of the species *k* and *j*

$$\Psi_{ACDB}^{(k,i,j)} = \Psi^{(i)} \tag{25}$$

The theory developed in this section is general for an arbitrary number of components, with an arbitrary number and functionality of association sites. The one restriction is that each molecule can have at most one pair of association sites which sterically hinder each other. In section III the theory is specialized to alcohols using the 3C-SH association model.

## III: Application to alcohols

In this section we specialize the theory developed in section II to the case of mixtures which contain alcohols with the 3C-SH association model outlined in Fig. 1. Site *H* (red) is the hydrogen bond donor site, $O_1$ is the first lone pair oxygen acceptor site (blue) and finally $O_2$ is the second lone pair oxygen acceptor site (green). Sites $O_1$ and $O_2$ are assumed equivalent and are sterically coupled. Hence, association at $O_1$ can block $O_2$ and vice versa. There are no steric effects between either oxygen site and the donor hydrogen *H*. Specializing Eqns. (12)-(13) to this case and enforcing Eq. (25)

$$c_{O_1}^{(k)} = c_{O_2}^{(k)} = \sum_{j=1}^{n} \sum_{B \in \Gamma^{(j)}} \rho^{(j)} X_B^{(j)} \Delta_{O_1 B}^{(k,j)} \tag{26}$$

$$c_H^{(k)} = \sum_{j=1}^{n} \sum_{B \in \Gamma^{(j)}} \rho^{(j)} X_B^{(j)} \Delta_{HB}^{(k,j)} + 2 \sum_{j=1}^{n} \sum_{i=1}^{n} \sum_{B \in \Gamma^{(j)}} \rho^{(i)} \rho^{(j)} X_B^{(j)} X_{O_1 O_2}^{(i)} \Delta_{HO_1}^{(k,i)} \Delta_{O_2 B}^{(i,j)} \left( \Psi^{(i)} - 1 \right) \tag{27}$$

$$c_{O_1 O_2}^{(k)} = \left( \sum_{j=1}^{n} \sum_{B \in \Gamma^{(j)}} \rho^{(j)} X_B^{(j)} \Delta_{BO_1}^{(j,k)} \right)^2 \left( \Psi^{(k)} - 1 \right) \tag{28}$$

$$c_{O_1 H}^{(k)} = c_{O_2 H}^{(k)} = 0 \tag{29}$$

Specializing Eqns. (15)-(16) to the 3C-SH model

$$X_{O_1}^{(k)} = X_{O_2}^{(k)} = X_{O_1 O_2}^{(k)} \left( 1 + \sum_{j=1}^{n} \sum_{B \in \Gamma^{(j)}} \rho^{(j)} X_B^{(j)} \Delta_{O_1 B}^{(k,j)} \right) \tag{30}$$

$$X_H^{(k)} = \frac{1}{1 + \sum_{j=1}^{n} \sum_{B \in \Gamma^{(j)}} \rho^{(j)} X_B^{(j)} \Delta_{HB}^{(k,j)} + 2 \sum_{j=1}^{n} \sum_{i=1}^{n} \sum_{A \in \Gamma^{(j)}} \rho^{(i)} \rho^{(j)} X_A^{(j)} X_{O_1 O_2}^{(i)} \Delta_{HO_1}^{(k,i)} \Delta_{O_2 A}^{(i,j)} \left( \Psi^{(i)} - 1 \right)} \tag{31}$$

$$X_{O_1O_2}^{(k)} = \frac{1}{\left(\sum_{j=1}^{n}\sum_{B\in\Gamma^{(j)}}\rho^{(j)}X_B^{(j)}\Delta_{O_1B}^{(k,j)}\right)^2\left(\Psi^{(k)}-1\right)+\left(1+\sum_{j=1}^{n}\sum_{B\in\Gamma^{(j)}}\rho^{(j)}X_B^{(j)}\Delta_{O_1B}^{(k,j)}\right)^2} \quad (32)$$

Combining Eqns. (30) and (32) we obtain

$$X_{O_1O_2}^{(k)}\left(\left(\frac{X_{O_1}^{(k)}}{X_{O_1O_2}^{(k)}}-1\right)^2\left(\Psi^{(k)}-1\right)+\left(\frac{X_{O_1}^{(k)}}{X_{O_1O_2}^{(k)}}\right)^2\right)=1 \quad (33)$$

Solving Eq. (33)

$$\delta^{(k)} = \frac{X_{O_1O_2}^{(k)}}{\left(X_{O_1}^{(k)}\right)^2} = \frac{2\Psi^{(k)}}{2\left(\Psi^{(k)}-1\right)+1+\sqrt{-4\left(\Psi^{(k)}-1\right)X_{O_1}^{(k)}\left(X_{O_1}^{(k)}-1\right)+1}} \quad (34)$$

Combining Eqns. (30) and (34)

$$X_{O_1}^{(k)} = X_{O_2}^{(k)} = \frac{1/\delta^{(k)}}{1+\sum_{j=1}^{n}\sum_{B\in\Gamma^{(j)}}\rho^{(j)}X_B^{(j)}\Delta_{O_1B}^{(k,j)}} \quad (35)$$

Hence the bonding fractions of molecules with the 3C-SH association model are obtained by the simultaneous solution of Eqns. (31) and (35), from which $X_{O_1O_2}^{(k)}$ can be calculated from Eq. (34).

For hydrogen bonding species in a multi-component mixture which do not exhibit steric hindrance, the fraction of molecules which are not bonded at the <u>acceptor</u> site $A_1$ is obtained from Eq. (35) by setting $O_1 = A_1$ and $\delta^{(k)} = 1$. Similarly, for hydrogen bonding molecules in a multi-component mixture which do not exhibit steric hindrance, the fraction of molecules not bonded at <u>donor</u> site $D_1$ is obtained from Eq. (31) with the transformation $H = D_1$.

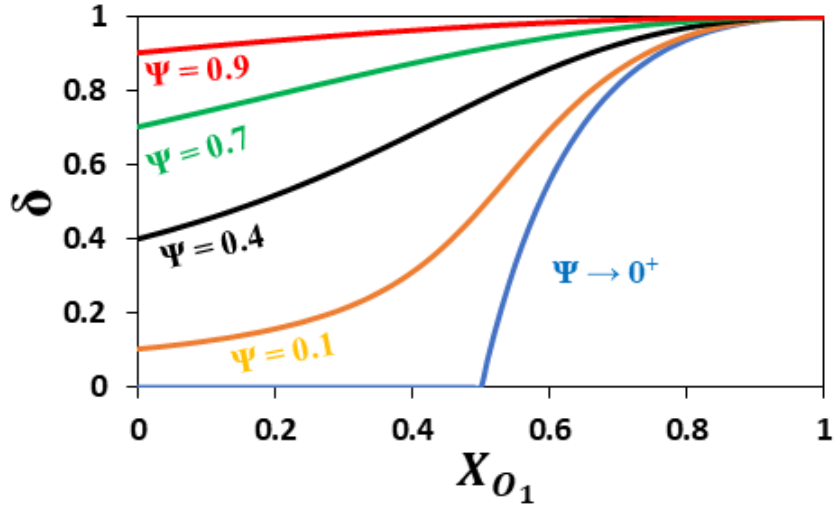

**Figure 2:** Plot of δ versus the fraction of unbonded oxygen sites (either $O_1$ or $O_2$) for various values of $\Psi$

Analyzing Eq. (34) we see that δ = 1 in the limit $\Psi$ = 1, signifying a reduction to TPT1 in the absence of steric hindrance. Figure 2 plots δ versus the fraction of unbonded oxygen sites (either $O_1$ or $O_2$) for various values of $\Psi$. The $\Psi \to 0^+$ curve is particularly interesting because δ vanishes identically for $X_{O_1}$ = 0.5. This demonstrates that the theory successfully reproduces complete blocking of the unbonded receptor site. When steric effects are included in models with a single acceptor and donor association sites, one must go to infinite order in perturbation to reproduce complete blockage.[29] For the current 3C-SH model with steric hindrance between $O_1$ and $O_2$, we obtain convergence at the TPT2 level.

## IV: Incorporation into simplified PC-SAFT

The association contribution alone is not sufficient to define the full equation of state. For this, we incorporate the association theory into the wider PC-SAFT[31] approach. In PC-SAFT molecules are modelled as chains of length $m$ of tangentially bonded hard spheres of diameter $\sigma$ with segment-segment square well attractions of depth $\varepsilon$.

The total excess (residual) Helmholtz free energy is defined as

$$a_{ex} = \frac{A_{ex}}{Nk_BT} = \bar{m}a_{hs} + a_{ch} + a_{at} + a_{as} \tag{36}$$

The association contribution to the free energy is described with the theory developed in this work Eq. (19). The contributions $a_{hs}$ and $a_{ch}$ are the excess free energy of the hard sphere reference fluid and the change in free energy associated with chain formation[12] from a fluid of hard spheres respectively. The average chain length in a mixture is given by (where $x_i$ is the mole fraction of species $i$)

$$\bar{m} = \sum_{i=1}^{n} x_i m_i \tag{37}$$

In this work we follow the simplified approach[16] of evaluating these contributions using the pure component Carnahan and Starling[32] forms of $a_{hs}$ and the contact value of the pair correlation function $g_{hs}$

$$a_{hs} = \frac{4\eta - 3\eta^2}{(1-\eta)^2} \quad ; \quad a_{ch} = (1-\bar{m})\ln g_{hs} \quad ; \quad g_{hs} = \frac{1-\frac{\eta}{2}}{(1-\eta)^3} \tag{38}$$

where the packing fraction is given by

$$\eta = \frac{\pi}{6}\sum_{i=1}^{n}\rho_i m_i d_i^3 \tag{39}$$

and $d_i$ is the temperature dependent hard sphere diameter

$$d_i = \sigma_i \left(1 - 0.12 \exp\left(-3 \frac{\varepsilon_i}{k_B T}\right)\right) \tag{40}$$

It has been demonstrated[21] that the use of Eqns. (38) instead of the standard mixture form[31] results in an equally capable equation of state. The use of the simplified approach in Eq. (38) is particularly useful in the current work, as it allows for a comparatively simple form of the association contribution to the chemical potential. This quantity is derived in Appendix A.

The contribution to the free energy due to isotropic attractions is given by $a_{at}$ in Eq. (36). Gross and Sadowski[31] developed $a_{at}$ using a modified Barker-Hendersen[33] second order perturbation theory (BH2) applied to chain molecules.

This completes the description of our TPT2 modification of PC-SAFT for the description of steric effects in 3C-SH alcohol models. In section V we apply the new theory to study both pure component and mixture phase equilibria of methanol and ethanol.

**V: Application to pure methanol and ethanol**

In this section we apply the new 3C-SH theory to study the phase equilibria and hydrogen bonding structure of pure methanol and ethanol. Alcohols are parameterized by 6 physically meaningful parameters: chain length $m$, hard sphere diameter $\sigma$, isotropic square well attraction energy $\varepsilon$, hydrogen bond volume $\kappa_{OH}$, hydrogen bond energy $\varepsilon_{OH}$ and the blockage integral $\Psi$. The standard parameterization approach for SAFT theories is to adjust the model parameters to vapor pressure ($P_{sat}$) and saturated liquid density data ($\rho_L$).[12] In this work, due to the need to obtain the blockage integral $\Psi$, we also include data for the heat of vaporization ($h_{vap}$). Specifically we adjust model parameters to minimize the objective function $J$ given as

$$J = \sum_{k=1}^{n_P} \frac{\left|P_{sat,data}^{(k)} - P_{sat,theory}^{(k)}\right|}{P_{sat,data}^{(k)}} + \sum_{k=1}^{n_\rho} \frac{\left|\rho_{L,data}^{(k)} - \rho_{L,theory}^{(k)}\right|}{\rho_{L,data}^{(k)}} + 0.2 \sum_{k=1}^{n_h} \frac{\left|h_{vap,data}^{(k)} - h_{vap,theory}^{(k)}\right|}{h_{vap,data}^{(k)}} \qquad (41)$$

where $n_p$, $n_\rho$ and $n_h$ is the number of vapor pressure, saturated liquid density, and heat of vaporization points respectively. Note, we have weighted the heat of vaporization data by a factor of 0.2.

We consider two cases in this study. For the first case we consider the 3C-SH model which treats alcohols with two oxygen lone pair acceptor and one hydrogen donor association sites where the two acceptor sites exhibit steric hindrance (TPT2). In the second case which we call TPT1, we treat the 3C association scheme in a first order perturbation theory which does not account for steric hindrance between the acceptor sites ($\Psi = 1$).

| Species | level | $m$ | $\sigma$ | $\varepsilon/k_B$ | $\varepsilon_{OH}/k_B$ | $\kappa_{OH}$ | $\Psi$ | AAD | | |
|---|---|---|---|---|---|---|---|---|---|---|
| | | | | | | | | $P_{sat}$ | $\rho_L$ | $h_{vap}$ |
| Methanol | TPT2 | 1.914 | 3.0361 | 206.565 | 2362.19 | 0.02898 | 0.2 | 0.70% | 1.71% | 3.87% |
| Methanol | TPT1 | 1.989 | 3.0123 | 224.729 | 2065.50 | 0.02699 | 1 | 0.28% | 2.57% | 6.57% |
| Ethanol | TPT2 | 2.746 | 3.0303 | 197.754 | 2294.91 | 0.02667 | 0.2 | 0.08% | 0.40% | 3.80% |
| Ethanol | TPT1 | 2.632 | 3.0978 | 214.965 | 2085.08 | 0.02065 | 1 | 0.06% | 1.20% | 5.60% |

**Table 1:** Model parameters for methanol and ethanol and average absolute deviations (AAD). The temperature ranges used in the data regression were $P_{sat}$ (240 K < T < 462 K), $\rho_L$ (160 K < T < 462 K) and $h_{vap}$ (300 K < T < 460 K). The parameters $\varepsilon/k_B$ and $\varepsilon_{OH}/k_B$ are in units K and $\sigma$ is in units of angstroms.

The resulting parameters and average absolute deviations can be found in Table 1. The best fit value for the blockage integral was found to be near $\Psi = 0.2$ for both alcohols. To force consistency, we then set the $\Psi = 0.2$ for both TPT2 parameter sets and refit the remaining parameters. A value of $\Psi = 0.2$ means that 80% of the allowable associated states to where both oxygen sites are bonded would be rejected due to overlap. Otherwise, the two TPT1 and TPT2 parameter sets are similar for each molecule. TPT2 gives better agreement with the *ab initio* calculated[34] hydrogen bonding energy of ethanol(zero point energy corrected MP2) $\varepsilon_{OH} = 2239.4$ $k_B$ K.

Both TPT2 and TPT1 accurately represent vapor pressure, liquid density and $h_{vap}$, although TPT2 gives a significant improvement in the representation of both liquid density and $h_{vap}$. However, TPT2 and TPT1 predict substantially different liquid phase hydrogen bond structure. Figure 3 compares model predictions to molecular dynamics simulations[28] using the OPLS-AA force field[35] for the fraction of ethanol molecules bonded $k$ times in a pure saturated liquid $\chi_k$. The fractions $\chi_k$ are derived in TPT2 in Appendix B.

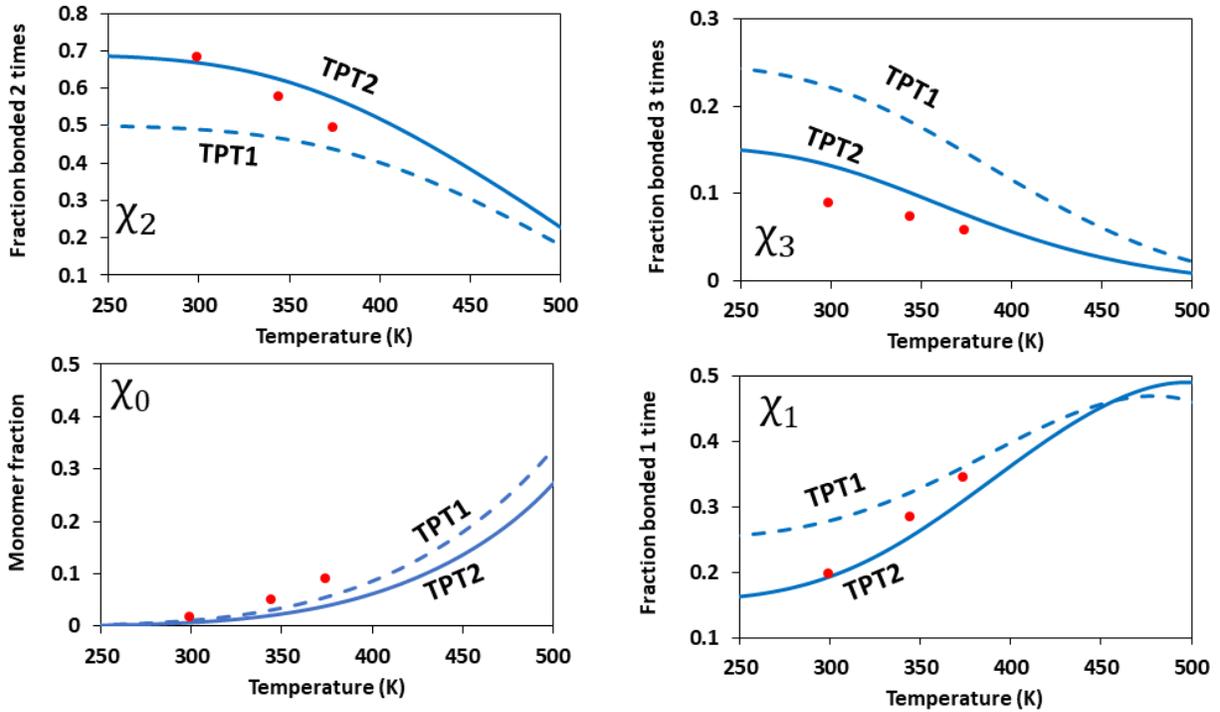

**Figure 3:** Comparison of MD simulations (circles)[28] and theory predictions: TPT2-solid curve, TPT1-dashed curve of the fraction of ethanol molecules bonded *k* times in a saturated liquid

We begin the discussion of Fig. 3 with a comparison of $\chi_2$ and $\chi_3$. As can be seen, TPT2 predicts values of $\chi_2$ and $\chi_3$ which are consistent with the MD simulation results, while TPT1 under-predicts $\chi_2$ and over-predicts $\chi_3$. This behavior is explicable by the fact that TPT1 does not account for steric effects between the two oxygen acceptor sites. Since there is no steric hindrance in TPT1, it overpredicts the fraction $\chi_3$. For a molecule to be fully bonded it must be bonded at both acceptor sites. As the sum $\sum \chi_k = 1$ must hold, the over-prediction $\chi_3$ must be debited from other fractions. This then results in the under-prediction of $\chi_2$ in TPT1. TPT2 is in better agreement with the simulated $\chi_1$ at lower temperatures while TPT1 appears to be in better agreement at high temperatures. Finally, both approaches under-predict the monomer fraction data (bottom left) as compared to both the MD simulations.

Please note, that the definition of a "hydrogen bond" will in general be different between simulation and theory. Hence, the comparison in Fig. 3 is qualitative in nature. Qualitatively, both TPT2 and MD predict a substantially lower $\chi_3$ than TPT1. However, for the T = 373 K data point, the comparison is not sufficiently quantitative to say whether TPT2 or TPT1 yields a more accurate prediction of $\chi_1$.

Table 2 compares model predictions to simulation data for the fractions $\chi_k$ of liquid methanol at $T$ = 300 K. The simulation results are taken from Ferrando et al.[36] who employed the AUA4 force field, which has been demonstrated to accurately represent liquid methanol structure. The TPT2 predictions accurately represent this simulation data, while TPT1 overpredicts the fraction $\chi_3$ and underpredicts $\chi_2$ for the same reasons discussed above for ethanol.

| Fraction | MC[36] | TPT2  | TPT1   |
|----------|--------|-------|--------|
| $\chi_0$ | 0.01   | 0.032 | 0.0064 |
| $\chi_1$ | 0.19   | 0.175 | 0.268  |
| $\chi_2$ | 0.67   | 0.68  | 0.493  |
| $\chi_3$ | 0.13   | 0.142 | 0.232  |

**Table 2:** Comparison of model predictions to Monte Carlo simulation predictions for the fraction of methanol molecules bonded $k$ times in a saturated liquid at T = 300 K.

**VI: Water-ethanol mixtures**

In section V it was demonstrated that TPT2 gave an improved representation of the hydrogen bonding structure of saturated liquid methanol and ethanol. In this section we focus on water-ethanol binary mixtures. However, before jumping into binary calculations, we first digress briefly to develop a benchmark water model.

### a. Development of a benchmark water model

PC-SAFT has found wide application as a general-purpose equation of state for multi-component phase equilibria; however, it has been demonstrated by several authors[37–40] that PC-SAFT does not give an entirely satisfactory representation the thermodynamics of pure water. The parameter sets for water which yield the best agreement with vapor pressures and saturated liquid densities, give a poor representation of the hydrogen bonded structure in the fluid. Recently, Marshall[40,41] demonstrated that if one accounts for the transition of water to tetrahedral symmetry within the second order Barker-Hendersen perturbation theory[2] for free energy $a_{at}$, an accurate equation of state for water could be obtained which accurately represents both the thermodynamics and hydrogen bonding structure of pure water.

In this work we develop a standard 4C (2 acceptor and 2 donor association sites) for water within TPT1, but we only develop the model for use over a limited temperature range for which it can act as an accurate "reference" when performing binary hydrogen bond structure and thermodynamics calculations with ethanol. When developing the water model, we include vapor pressure and liquid density data in the temperature range 293 K ≤ T ≤ 473 K. In addition, to optimize the choice of hydrogen bonding parameters, we include hydrogen bond structure data on the fraction of free OH groups ($X_A$) as measured by Luck[42] in the data regression over this same temperature range. We adjust model parameters to minimize the following objective function

$$J = \sum_{k=1}^{n_P} \frac{\left|P_{sat,data}^{(k)} - P_{sat,theory}^{(k)}\right|}{P_{sat,data}^{(k)}} + \sum_{k=1}^{n_\rho} \frac{\left|\rho_{L,data}^{(k)} - \rho_{L,theory}^{(k)}\right|}{\rho_{L,data}^{(k)}} + \sum_{k=1}^{n_X} \frac{\left|X_{A,data}^{(k)} - X_{A,theory}^{(k)}\right|}{X_{A,data}^{(k)}} \tag{42}$$

where $n_X$ is the number of $X_A$ data points. The results can be found in Table 3. Over this limited temperature range, PC-SAFT can accurately represent vapor pressure, saturated liquid density and the fraction of free OH groups.

| m | $\sigma$ (Å) | $\varepsilon/k_b$ (K) | $\varepsilon_{OH}/k_b$ (K) | $\kappa_{OH}$ | AAD $P_{sat}$ | $\rho_L$ | $X_A$ |
|---|---|---|---|---|---|---|---|
| 1 | 3.063 | 350.624 | 1502.33 | 0.031196 | 0.28% | 1.50% | 2.10% |

**Table 3:** Regression results for 4C water model in the temperature range 293 K ≤ T ≤ 473 K

### b. Predictions of Ethanol-water hydrogen bonding

For the ethanol-water pair, the cross-association bond volume and association energy are given by the combining rules in Eq. (21). There is no binary parameter in the association contribution of the theory to tune to experimental data. Hence any calculation of binary hydrogen bonding structure is necessarily a prediction. In Fig. 4 we compare TPT1 and TPT2 predictions to MD simulations (same reference[28] and methodology as Fig. 3 for ethanol, with water modeled with the *i*-AMOEBA[43] force field) for the average number of hydrogen bonds per ethanol molecule $n_{HB}(x)$, where $x$ is the mole fraction of ethanol, in an ethanol-water binary liquid mixture at T = 298 and 373 K. As can be seen, the $n_{HB}(x)$ composition dependence of TPT2 is in much better agreement with simulation than TPT1. Most notably, TPT2 predicts much less hydrogen bonding in the ethanol dilute region, due to the inclusion of steric effects.

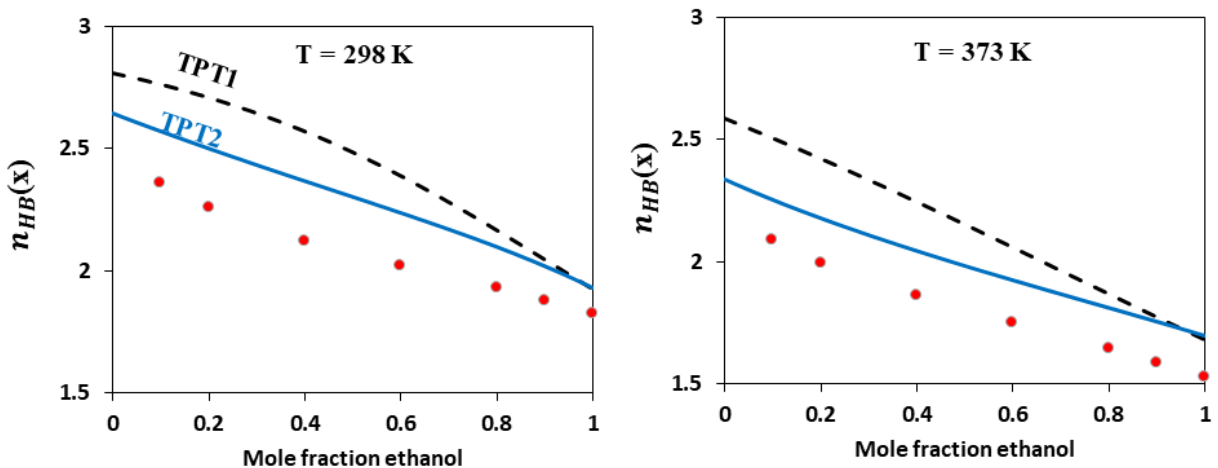

**Figure 4:** Comparison of model predictions (dashed curve-TPT1, solid curve-TPT2) to MD simulation results (circles)[28] for the average number of bonds per ethanol molecule in a water/ethanol binary mixture at T = 298 K (left) and T = 373 K (right)

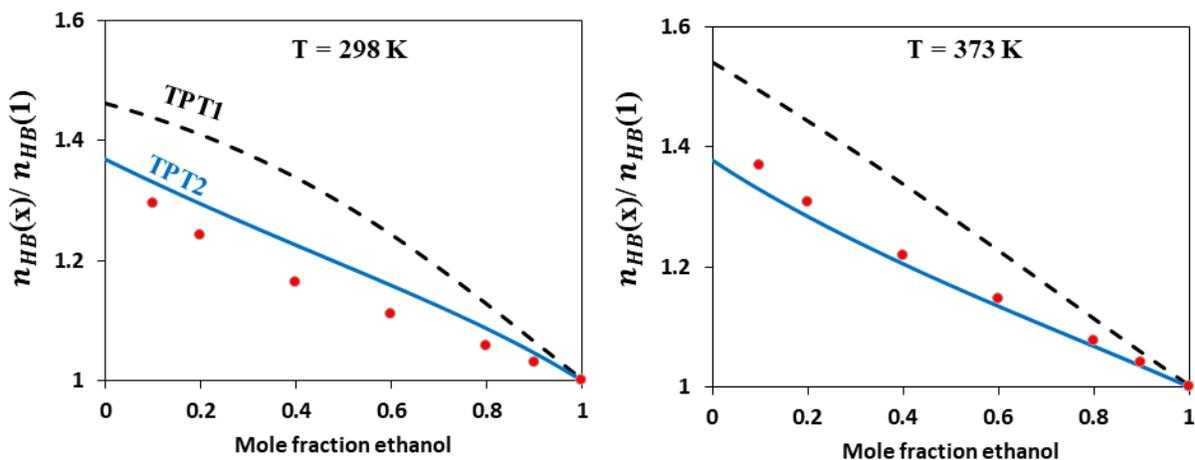

**Figure 5:** Same as Fig. 4, except for all results scaled by the average number of hydrogen bonds in pure ethanol $n_{HB}(1)$

The MD calculation of $n_{HB}(x)$ in Fig. 4 is ambiguous in the sense that one must impose on the simulation the definition of what it means to be hydrogen bonded. Hence, the absolute value of the simulated $n_{HB}(x)$ is less telling than the composition dependence. In Fig. 5 we present the same results as Fig. 4, but we scale all results by the average number of hydrogen bonds in pure ethanol $n_{HB}(1)$. As can be seen, the composition dependence of TPT2 is in good agreement with the simulation results, while TPT1 clearly shows a stronger composition dependence.

### c. Ethanol-water phase equilibria and heat of mixing

To describe the binary phase equilibria and heat of mixing for the ethanol-water binary system we must first adjust the binary interaction parameter $k_{ij}$; which is used in the calculation of the cross square well depth $\varepsilon_{ij}$ in the free energy term due to isotropic attractions $a_{at}$. We use standard Lorentz-Berthelot mixing rules

$$\varepsilon_{ij} = \sqrt{\varepsilon_i \varepsilon_j}\left(1-k_{ij}\right) \tag{43}$$

The binary parameter $k_{ij}$ is adjusted to minimize the error in the description of binary vapor-liquid equilibria (VLE).

| Method | $k_{ij}$ |
|--------|----------|
| TPT1   | -0.02424 |
| TPT2   | -0.06413 |

**Table 4:** Binary parameters for the ethanol-water pair

Table 4 gives regressed values of $k_{ij}$ using both TPT1 and TPT2, and Fig. 6 compares TPT1 and TPT2 model results to experimental data for the phase diagram at atmospheric pressure. As can be seen, both TPT1 and TPT2 are able to accurately correlate the binary VLE data, although TPT2 does give an improved result.

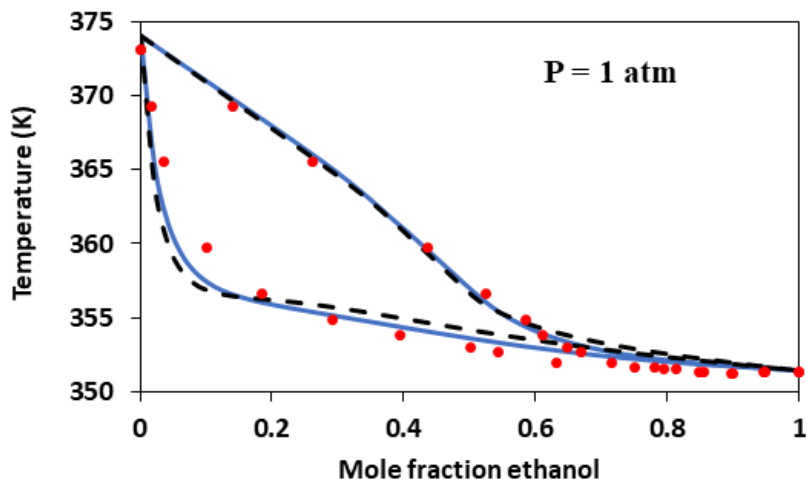

**Figure 6:** Binary phase equilibria model results (TPT1-dashed cure, TPT2-solid curve) compared to experimental data (circles)[44] for ethanol-water binary at atmospheric pressure

While both TPT1 and TPT2 give a similar correlation of binary VLE, TPT2 is substantially more accurate in the prediction of the heat of mixing $h_{mix}$. This can be seen in Fig. 7 which compares model predictions of $h_{mix}$ to experimental data at several temperatures. While neither approach is accurate at the T = 323.15 K, TPT2 gives much more accurate predictions of $h_{mix}$ at the higher temperatures T = 373.15 K and T = 398.15 K. For low ethanol concentrations, both TPT1 and TPT2 give similar predictions for $h_{mix}$. However, for ethanol rich mixtures, TPT1 substantially under-predicts $h_{mix}$, while TPT2 is in good agreement with the experimental data. This is due to the inclusion of steric effects in the TPT2 theory.

The disagreement between model and experiment at T = 323.15 K occurs at a temperature which coincides with a change from $h_{mix} > 0$ to $h_{mix} < 0$ as temperature is decreased. It is also in this temperature regime that water is undergoing a structural transition towards tetrahedral symmetry.[45] This transition is not included in a standard PC-SAFT model for water. Hence, the disagreement at T = 323.15 K for $h_{mix}$ is likely the result of the PC-SAFT water model and not of the theoretical treatment of ethanol itself. Recently, Marshall[40] included water's transition to tetrahedral symmetry in the PC-SAFT equation of state by means of an associated reference fluid. A point of future research could be to combine the approach of Marshall, which accounts for the transition to tetrahedral symmetry of fully hydrogen bonded water, with the TPT2 alcohol model developed in this work.

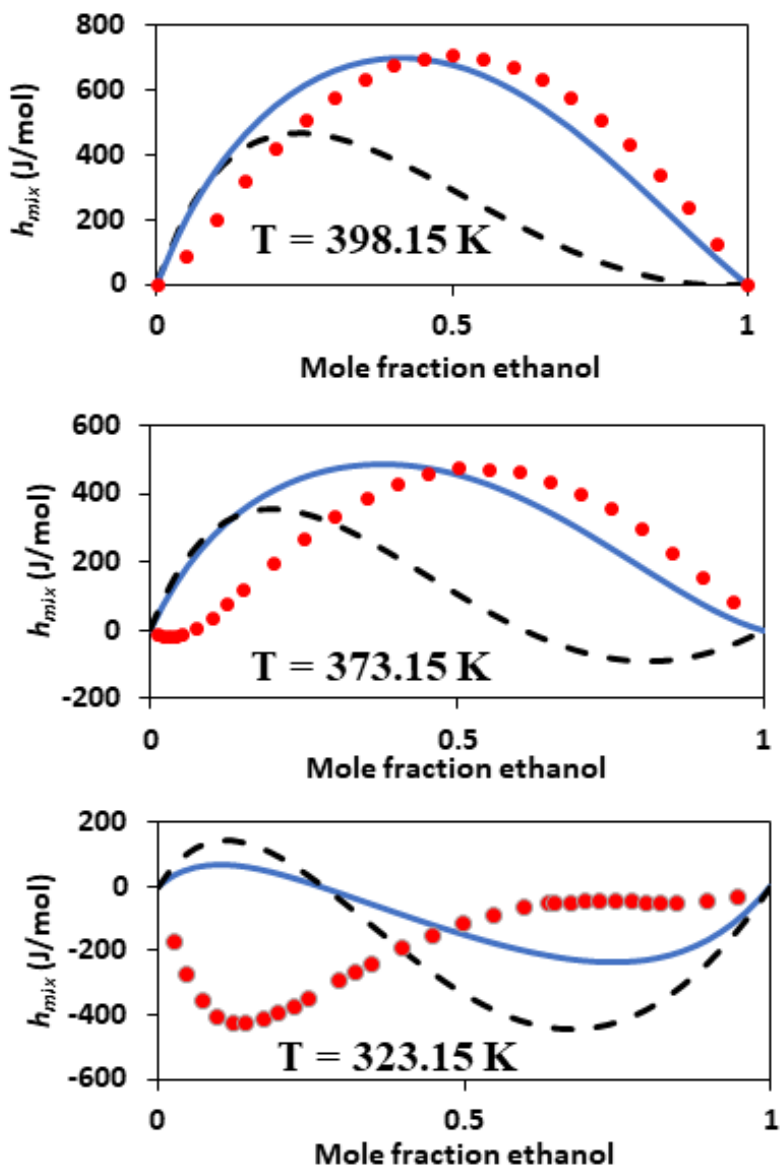

**Figure 7:** Heat of mixing (change of enthalpy of mixing) as a function of ethanol mole fraction in a water ethanol mixture. Solid curve give TPT2 predictions, dashed curve TPT1 predictions and symbols give experimental data[46,47]

When inserting a water molecule into an ethanol rich liquid phase, TPT1 will predict overly accessible oxygen acceptor sites on the ethanol molecule, while TPT2 correctly accounts for the fact that when one ethanol acceptor site is occupied, it partially blocks the second. This results in an increase in $h_{mix}$ as compared to TPT1. Figure 9 demonstrates this hydrogen bonding effect by showing the theory predicted average number of hydrogen bonds per water molecule in an ethanol / water binary mixture at a temperature of 373.15 K. As the mole fraction of ethanol increases, so does the deviation between TPT1 and TPT2. When water is dilute, TPT2 predicts substantially less water hydrogen bonding than TPT1.

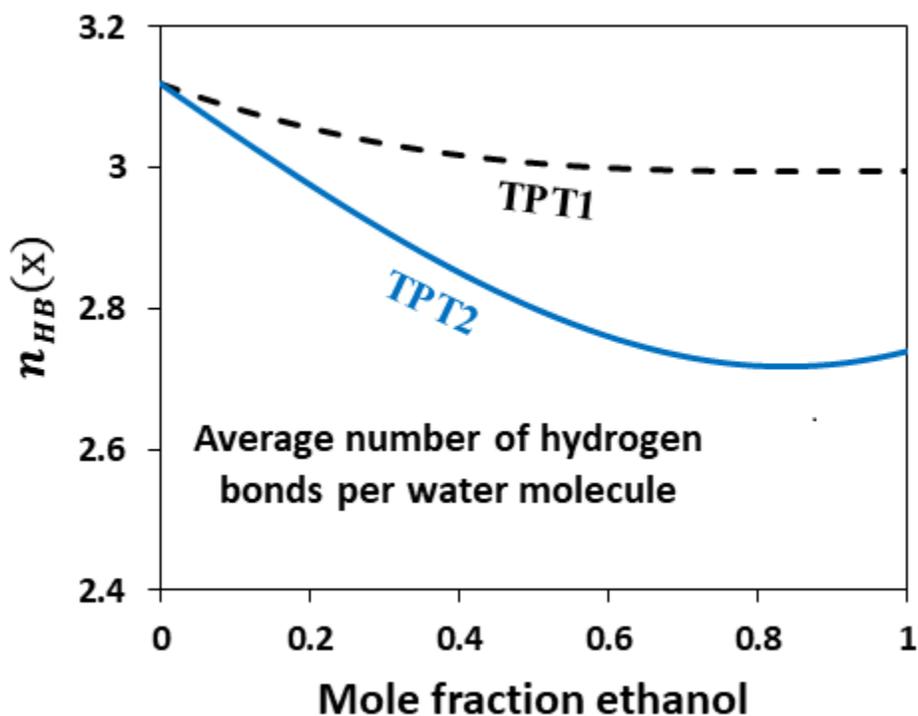

**Figure 8:** Theory predictions of the average number of liquid phase hydrogen bonds per water molecule in an ethanol / water binary mixture at T = 373.15 K

## VI: Conclusions

In this paper we have extended TPT2 to account for steric effects in associating molecules. The theory is derived for a multi-component fluid where each molecule can have an arbitrary number of acceptor and donor association sites. However, for mathematical tractability, we have restricted the theory such that only a single pair of association sites are allowed to sterically hinder each other. The theory was then applied to a 3C-SH alcohol model where steric hindrance between the two oxygen acceptor sites was accounted for in the model. It was demonstrated in Fig. 2 that complete blockage could be achieved at the TPT2 level. We then paired this new association theory with the PC-SAFT equation of state to study hydrogen bonding and phase equilibria of alcohols. It was demonstrated that accounting for steric hindrance in TPT2 allows for a more capable equation of state as compared to a traditional first order approach.

**Appendix A: Derivation of the association contribution to the chemical potential**

In this appendix we derive the association contribution to the chemical potential in Wertheim's multi-density formalism. In general, the association contribution to the chemical potential is given by the relation[9]

$$\frac{\mu_{as}^{(k)}}{k_b T} = \ln X_o^{(k)} - \frac{\partial \Delta c^{(o)}/V}{\partial \rho^{(k)}} \tag{44}$$

From Eqns. (5)-(8)

$$\frac{\partial \Delta c^{(o)}/V}{\partial \rho^{(k)}} = \frac{1}{2} \frac{\partial \ln g_{hs}}{\partial \rho^{(k)}} \sum_{i=1}^{n} \rho^{(i)} \sum_{A \in \Gamma^{(i)}} X_A^{(i)} c_A^{(i)} + F^{(k)} \tag{45}$$

Specializing to the 3C-SH association model

$$F^{(k)} = \frac{\partial \ln g_{hs}}{\partial \rho^{(k)}} \sum_{i=1}^{n} \left( \sum_{j=1}^{n} \sum_{B \in \Gamma^{(j)}} \rho^{(j)} X_A^{(j)} \Delta_{AO_1}^{(j,i)} \right)^2 \rho^{(i)} X_{O_1O_2}^{(i)} \left( \Psi^{(i)} - 1 \right)$$

$$= \frac{\partial \ln g_{hs}}{\partial \rho^{(k)}} \sum_{i=1}^{n} \rho^{(i)} X_{O_1O_2}^{(i)} \left( \frac{X_{O_1}^{(i)}}{X_{O_1O_2}^{(i)}} - 1 \right)^2 \left( \Psi^{(i)} - 1 \right)$$

(46)

Combining Eqns. (15), (45) and (46)

(47)

$$\frac{\partial \Delta c^{(o)}/V}{\partial \rho^{(k)}} = \frac{\partial \ln g_{hs}}{\partial \rho^{(k)}} \left( \sum_{i=1}^{n} \rho^{(i)} \left( \frac{1}{2} \sum_{a \in \Gamma^{(i)}} \left( 1 - X_A^{(i)} \right) \right) + \sum_{i=1}^{n} \rho^{(i)} \left[ \frac{\left(X_{O_1}^{(i)}\right)^2}{X_{O_1O_2}^{(i)}} - 1 + X_{O_1O_2}^{(i)} \left( \frac{X_{O_1}^{(i)}}{X_{O_1O_2}^{(i)}} - 1 \right)^2 \left( \Psi^{(i)} - 1 \right) \right] \right)$$

Where the second sum on the right-hand side of Eq. (A4) is over the species of 3C-SH molecules only.

### Appendix B: Derivation of the fraction of molecules bonded $k$ times $\chi_k$

In Wertheim's multi-density statistical mechanics, the density of molecules bonded at the set of sites $\gamma$ is given by[9]

$$\frac{\rho_\gamma^{(k)}}{\rho_o^{(k)}} = \sum_{P(\gamma)=\{\tau\}} \prod_\tau c_\tau^{(k)}$$

(48)

where $P(\gamma)$ is the partition of the set $\gamma$ into non-empty subsets. For the 3C-SH association model, the density of molecules bonded once at any association site is given by

$$\frac{\rho_1^{(k)}}{\rho_o^{(k)}} = 2c_{O_1}^{(k)} + c_H^{(k)}$$

(49)

The density bonded twice

$$\frac{\rho_2^{(k)}}{\rho_o^{(k)}} = \left(c_{O_1}^{(k)}\right)^2 + 2c_H^{(k)} c_{O_1}^{(k)} + c_{O_1 O_2}^{(k)} \tag{50}$$

And the density bonded three times

$$\frac{\rho_3^{(k)}}{\rho_o^{(k)}} = \left(c_{O_1}^{(k)}\right)^2 c_H^{(k)} + c_{O_1 O_2}^{(k)} c_H^{(k)} \tag{51}$$

The fraction of molecules bonded $j$ times are calculated through the relation

$$\chi_j^{(k)} = \frac{\rho_j^{(k)}}{\rho^{(k)}} \tag{52}$$

[47] J.B. Ott, G.V. Cornett, C.E. Stouffer, B.F. Woodfield, C. Guanquan, and J.J. Christensen, J. Chem. Thermodyn. **18**, 867 (1986).